\documentclass[12pt,preprint]{aastex}

  \usepackage{natbib}
  \bibpunct{(}{)}{;}{a}{}{,} % for A&A style         

\newcommand{\fluxunits}{10$^{-16}$ erg~s$^{-1}$~cm$^{-2}$~\AA$^{-1}$}

\shorttitle{Optical Spectrum of the Radio Jet 3C 120}
\shortauthors{S\'anchez et al.}

\begin{document}

%\title{Decoupled optical spectrum of the 4$\arcsec$ knot of the radio jet of
%3C 120}
\title{Clean optical spectrum of the radio jet of 3C 120}

\author{S.F~S\'anchez\altaffilmark{1},
  B.Garcia-Lorenzo\altaffilmark{2}, E.Mediavilla\altaffilmark{2},
  J.I. Gonz\'alez-Serrano\altaffilmark{3,4}, L.Christensen\altaffilmark{1}}

\altaffiltext{1}{Astrophysikalisches Institut Potsdam, An der Sternwarte 16, 14482 Potsdam, Germany}
\email{ssanchez@aip.de}
\altaffiltext{2}{Instituto de Astrof\'\i sica de Canarias, 38205 La Laguna,
  Tenerife, Spain}
\altaffiltext{3}{Instituto de Fisica de Cantabria, UC-CSIC, Av. de Los Castros
  S/N, 35005, Santander, Spain}
\altaffiltext{4}{Dept. de Fisica Moderna, Facultad de Ciencias, UC, Av. de Los Castros
  S/N, 35005, Santander, Spain}

\begin{abstract}
  
  We present integral field spectroscopy (IFS) of the central region of
  3C~120. We have modeled the nuclear and host galaxy 3D spectra using
  techniques normally applied to imaging, decoupling both components, and
  obtained a residual datacube. Using this residual datacube, we detected the
  extended emission line region associated with
%4$\arcsec$ knot of 
  the radio jet.  We obtained, for the first time, a clean spectrum of
  this region and found compelling evidences of a jet-cloud
  interaction. The jet compresses and splits the gas cloud which is
  ionized by the AGN and/or by the strong local UV photon field
  generated by a shock process.  We cannot confirm the detection of an
  extended emission line region associated with the counter-jet
  reported by \cite{axon89}.

%Using it we found compelling evidences of a jet-cloud interaction,
%  although we could not clarify if the dominant ionization mechanism is
%  photoionization by the AGN or by the strong local UV photon field generated
%  by a shock process.

\end{abstract}


\section{Introduction}

3C~120 is a Seyfert 1 radio-galaxy at a redshift of 0.033 with a compact
flat-spectrum radio core and a powerful X-ray emission \citep{halp85}. It has
a one-side superluminal radio jet with two clear knots at 4$\arcsec$ west and
25$\arcsec$ north-west from the core \citep{walk87,walk88,walk97}.
Morphologically it is a bulge-dominated galaxy with several continuum and
extended emission line regions
\citep[EELR,][]{bald80,wler81,pere86,hua88,soub89}, indicating a change of
star formation history and a combination of different ionization mechanisms
from region to region.

The radio-jet shows a continuum dominated optical counterpart, consistent with
synchrotron radiation, that extends from 7$\arcsec$ to 25$\arcsec$ out of the
nucleus, but without a clear counterpart at the location of the 4$\arcsec$
knot \citep{hjor95}. X-ray emission has been detected associated with the
25$\arcsec$ knot \citep{harr99}.  An EELR has been detected at
$\sim$1$\arcsec$ distance from the 4$\arcsec$ knot, the so called E$_1$ region
\citep{soub89}. This EELR has been detected using different spectroscopic and
imaging techniques \citep[e.g.][]{bald80,wler81,hua88,soub89,axon89}.
\cite{axon89} showed that it comprises two kinematic components, separated by
$\sim$150 km~s$^{-1}$, with a South-North gradient.
%The north component is
%highly polarized, while the south component is not \citep{soub89}. 
%This may
%indicate that the north component is behind the jet (and gets polarized while
%crossing it), and the south component is in front of it. 
This EELR does not originate in material that has been ejected with the
radio-emitting plasma. It arises because of the interaction, most probably
lateral expansion, of the radio jet with the interstellar medium (ISM) of the
galaxy \citep{axon89}.  Very little is known about the ionization conditions
or the amount of dust in this area.  \cite{hua88} estimated roughly an
[OIII]$\lambda$5007/H$\beta$ ratio of $\sim$17 for this cloud, based on the
data from \cite{bald80}, which contrasts with the value found for the nucleus
($\sim$1). These values were derived without a proper deblending of the broad
and narrow emission lines, and without decontaminating the spectrum of the
E$_1$ region from the contribution of the host galaxy.
%Similar caveants may be
%applied to the data presented by \cite{axon89}.

\section{Observations and Data Reduction}

We obtained integral field spectroscopy (IFS) of the core of 3C~120 using
INTEGRAL \citep{arr98} at the WHT telescope to derive a clean spectrum of the
E$_1$ region.  The observations were done on the 26th of February 2003, under
photometric conditions and an average seeing disk of 1.2$\arcsec$ FWHM. We
obtained 3 exposures of 1200s using the SB2 fiber bundle, which comprises 219
fibers (189 science+30 sky fibers) each one of 0.9$\arcsec$ in diameter
projected on the sky, covering a total field-of-view of
$\sim$16$\arcsec$$\times$12$\arcsec$.  We used the low-resolution V300 grating
centered at $\sim$5500\AA, with a nominal spectral resolution of $\sim$450 at
the central wavelength, covering the range between 3500-9000\AA\ with a final
sampling of $\sim$3\AA/pixel. A detailed explanation of the data reduction and
calibration will be presented elsewhere (Garcia-Lorenzo et al.  in prep.). It
basically comprises the standard steps, including bias subtraction, spectra
extraction, flat-fielding, wavelength and flux calibration. Finally, an
interpolation routine was applied over each monochromatic slice using E3D
\citep{sanc04}, to obtain a regular grid datacube with 0.3$\square\arcsec$
pixel.  For comparison purposes we obtained a WFC F555W-band image of the
3C~120, available in the archive of the Hubble Space Telescope (HST). This
image has a 3$\sigma$ limiting surface brightness magnitude of $\sim$25.5
mag/arcsec$^{2}$, deep enough for the proposes of this article.
%The
%contamination of flux coming from the emission lines is negligible in
%the F555W-band image. Therefore, 
Due to the width of the F555W-band this image traces the morphology of the
continuum emission in 3C~120.

\section{Analysis and Results}

\begin{figure}
    \resizebox{\hsize}{!}  {\includegraphics[width=\hsize]{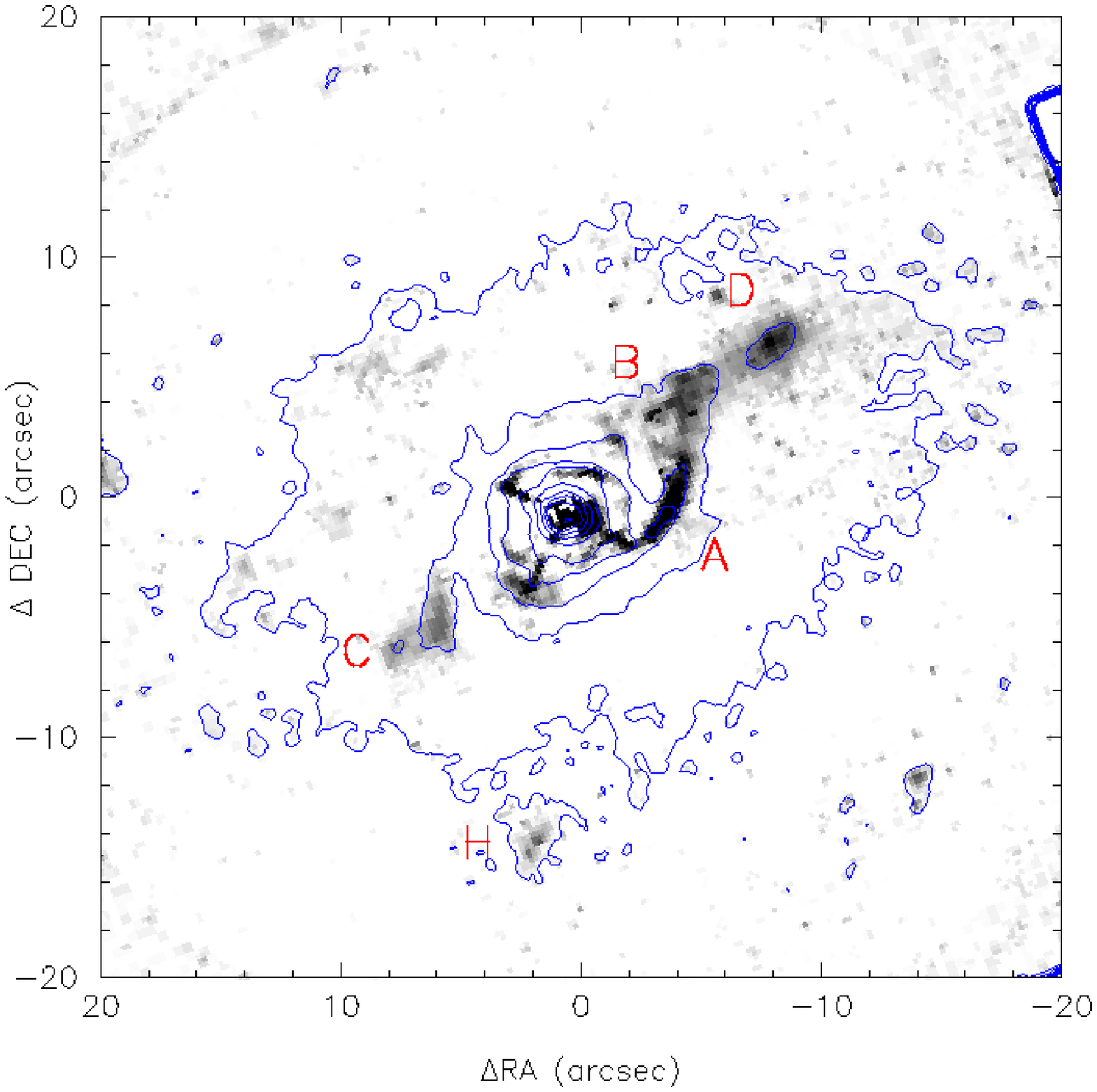}}
  \caption{\label{hst} 
    Contour-plot of the F555W-band image of 3C 120 obtained with the
    HST/WFPC2. The grey-scale shows the residuals after subtracting the smooth
    component by an isophotal analysis. The spikes of the WFC PSF are clearly
    seeing. We have marked the position of the continuum dominated
    substructures using the nomenclature of \cite{soub89}.
  }
  \end{figure}

We create a model from the F555W-band image of the object using an isophotal
surface brightness analysis, following the prescriptions described by
\cite{jedr87}.
% implemented in our own coded software. 
%This method provides us
%with a model of the object without any other assumption but an elliptical
%shape for the isophotes. 
The model was subtracted from the original image,
obtaining a residual image. 
%This residual image shows the substructures present in
%the continuum of 3C~120 \cite[e.g.][]{soub89}. 
\cite{hjor95} and \cite{soub89} used this technique to detect the optical
counterpart of the radio jet in 3C~120 and the EELRs, respectively.  Figure
\ref{hst} shows the resulting residual image in an arbitrary greyscale
together with the contours of the original image. The different
continuum-dominated substructures detected by \cite{soub89} are indicated in
the figure using their nomenclature. As expected, there is no trace of the
emission-line dominated substructures.

\begin{figure*}[p]
\resizebox{\hsize}{!}
%{\includegraphics[width=\hsize]{maps_art6_gimp.ps}}
{\includegraphics[width=\hsize]{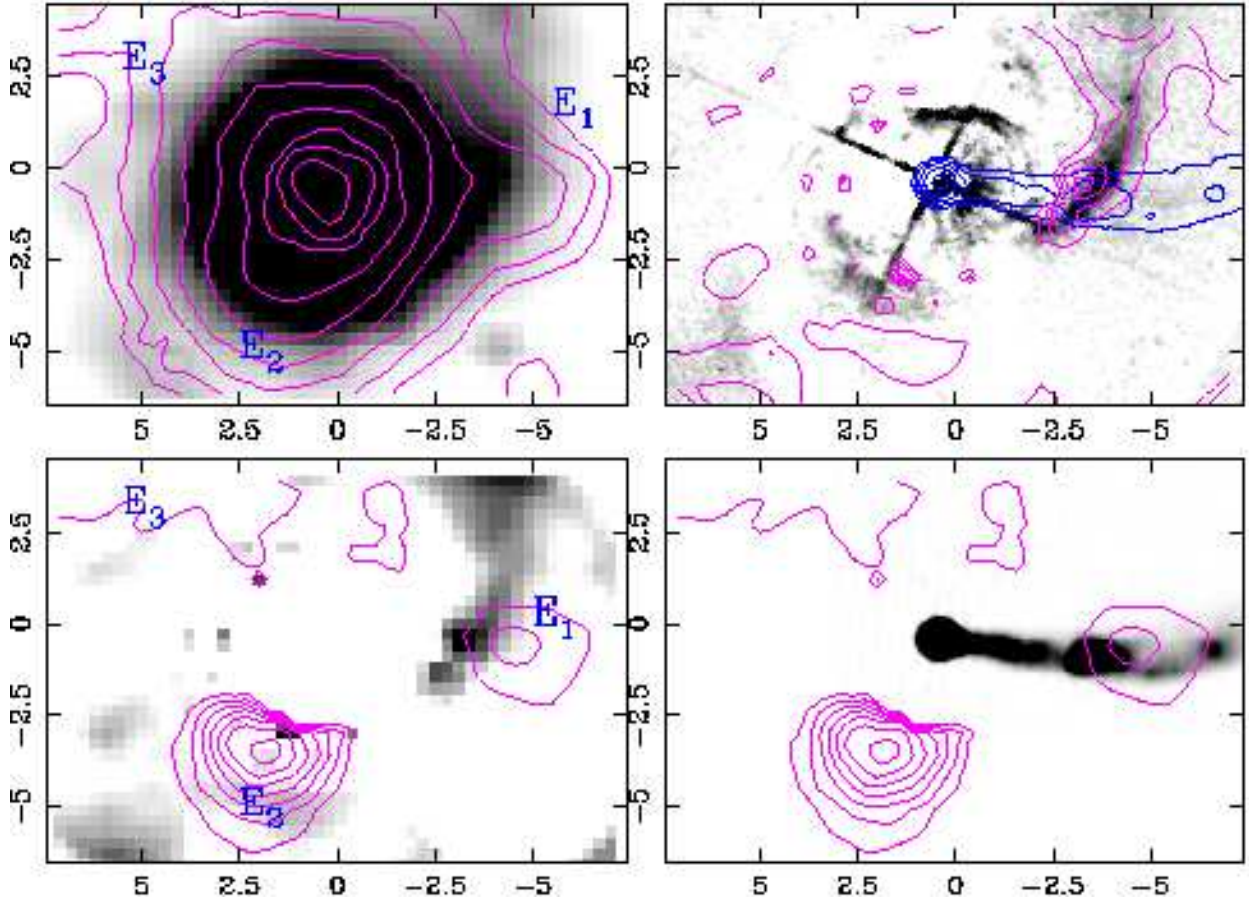}}
%{\includegraphics[width=\hsize]{maps_art6.ps}}
  \caption{\label{maps} 
    Top-left: contour-plot of a narrow-band image centred on the
    [OIII]$\lambda$5007 emission line at the redshift of the object
    (5170-5200\AA) together with a
    greyscale narrow-band image of the adjacent continuum (5204-5246\AA)
    obtained from the original datacube. The continuum emission was
    subtracted from the emission-line map.  The contours start at
    0.5$\times$\fluxunits, with 0.25 dex separation between each one. E$_1$,
    E$_2$ and E$_3$ indicate the [OIII] emission areas with the nomenclature
    of Soubeyaran et al.  (1989).  Top-right: contour-plot of a narrow-band
    image centred on the continuum adjacent to the [OIII] emission line
    (5204-5246\AA) obtained from the residual datacube, together with a
    greyscale of the residual from the HST F555W broad-band image (Fig.
    \ref{hst}). The contours start at 0.02$\times$\fluxunits, with a
    separation of 0.015$\times$\fluxunits.  The blue contours show the map of
    the radio jet at 4885 MHz.  Bottom-Left: contour-plot of a narrow-band
    image centred on the [OIII]$\lambda$5007 emission line (5170-5200\AA)
    together with a greyscale of the adjacent continuum (5204-5246\AA)
    obtained from the residual datacube.  The contours start at
    0.5$\times$\fluxunits, with a separation of 0.3$\times$\fluxunits.
    Bottom-Right: the same contour-plot as that in the bottom-left figure,
    together with a greyscale of the map of the radio jet at 4885 MHz.  We
    cannot confirm the detection of an EELR associated with the undetected
    counter-jet, reported by \cite{axon89}, at $\sim$5$\arcsec$ east from the
    nucleus.
  }
  \end{figure*}

Figure \ref{maps}, top-left panel, shows the intensity map of the
[OIII]$\lambda$5007 emission line (contours) and the adjacent continuum
(greyscale). These maps have been created coadding the intensities of the
5170-5200\AA\ and 5204-5246\AA\ wavelength ranges in the datacube,
respectively, and subtracting the continuum from the emission line map.  The
emission line map is remarkably similar to the narrow-band image presented by
\cite{hua88}, showing a strong gaseous emission extended through out the host
galaxy.  Despite the strong contamination from the nucleus and the hosts
galaxy, it is possible to identify the [OIII] EELRs described by \cite{soub89}
(marked as E$_1$, E$_2$ and E$_3$). To derive clean spectra of these
substructures we create a galaxy+nucleus 3D template datacube and subtract it
from the data.

First, we create for each wavelength a narrow band image of the width of one
spectral pixel ($\sim 3$\AA) from the original datacube. Then, we model the
object at each of these {\it monochromatic} images using two different
techniques widely used the analysis of 2D images: (1) a 2D image modelling of
the nucleus and the host and (2) an isophotal modelling based on a surface
brightness analysis. The first method has the advantage that it also provides
us with a decoupled spectra of the nucleus and the host galaxy \cite[see
][]{jahn04}. Previous attempts of doing so have used long-slit spectrocopy,
with the inherent limitations due to the loss of the 2D information
\citep{jahn02}. A similar technique has been used successfully for the
deblending of QSO lenses with IFS \citep{wiso03}. However, this method has the
disadvantage that it requires a good estimation of the PSF, and it requires
more assumptions about the morphology of the object (i.e., more parameters to
fit).  A more detailed description of the technique and its limitations will
be presented elsewhere (S\'anchez et al., in prep.).

The 2D image modelling of the nucleus and the host for each {\it
  monochromatic} image is performed using GALFIT \citep{peng02}. This program
has been extensively tested in the image decomposition of QSO/hosts
\citep{sanc04b}. The 2D model comprises a narrow gaussian function (to model
the nucleus) and a de~Vaucouleurs law (to model the galaxy) both convolved
with a PSF. The PSF was obtained from a calibration star datacube, observed
just before the object. The structural parameters of the host galaxy (PA,
ellipticity and effective radius) were obtained by a 2D modelling of the
F555W-band image. The fit over the datacube was performed twice: (1) leaving
the centroid of the object and the intensities of both components free, and,
(2) fixing the centroid by the result of a polynomial fitting as a function of
the wavelength over the results of the first fit.
%Therefore, only the intensities of both components were fitted this second
%time. 
This increases the accuracy of the recovered spectra
%model by the use of the 3D information and
%the high quality HST data 
\cite[see ][]{wiso03}. The final 3D datacube model
is then subtracted from the original datacube to get a residual datacube.
  
As a second method, we perform a surface brightness analysis of each
monochromatic image, using the same code that we used for the analysis of the
F555W-band image (Fig. \ref{hst}). This method provides us with a 2D model of
the object, which we subtract from the original image to obtain a residual
image for each wavelength. Like in the previous method, the technique was
applied twice: (1) once to let the program look for the best elliptical
isophote at each radii. The centroid, the PA, the ellipticity and the
intensity are the free parameters; and (2), fixing all the parameters but the
intensity to the values derived from a polynomial fitting as a function of the
wavelength over the results of the first pass. We subtract the final 3D model
from the original data to obtain a residual datacube.

The residual datacubes obtained using both techniques are quite similar.
However, the residual produced by the first method shows a ring structure in
the inner region at any wavelength. This structure, $\sim$1000 times fainter
than the peak intensity of the nucleus, is not seen in the residual of the
F555W-band image. Similar structures are normally found in this kind of 2D
modelling due to inaccuracies in the determination of the PSF. This was our
case, since we have clearly undersampled the PSF.
%: each fiber has a diameter of
%0.9$\arcsec$ am
Although it does not strongly affect the extracted spectra of the nucleus and
the host, it introduces a non poissonian noise in the residual datacube. Due
to that we restricted our analysis of the residuals to the datacube obtained
with the second method, using the first method only for deblending the host
and nucleus contributions. 

It is important to note here that the PSF undersampling does not affect the
accuracy of the centroid determination, fundamental to compare images taken
with different instruments. The displacement of the object along the
field-of-view at different wavelengths due to differential atmospheric
refraction \citep{fili82} can be used to determine the centroid position with
high precision \citep{medi98,arr99,wiso03}. The accuracy in the determination
of the centroid in our IFS was better than 0.1$\arcsec$, at any wavelength
(S\'anchez et al., in prep.). 
%Similar results were found by \cite{medi98} and
%\cite{wiso03}.

%  and to cross-check the results.

Figure \ref{maps}, top-right panel, shows a contour-plot of the narrow-band
image at the continuum adjacent to the [OIII]$\lambda$5007 emission lines
(5204-5246\AA) extracted from the residual datacube. The grey-scale shows the
residual of the continuum dominated F555W-band image. Both images were
recentered by matching the peak of the central point-like source in the
F555W-band image with the centroid of the object in the IFS data. The accuracy
of the determination of the position of the peak in the HST image is a
fraction of the pixel. Thus, the error in the recentering was dominated by the
error in the determination of the centroid in the IFS data ($<0.1\arcsec$, as
quoted above). We used the rotator angle of the WHT and the WCS of the HST
image to align both images in the sky. Despite the superior resolution of the
HST image, the agreement between both maps is remarkable.  There is an
expected missmatch in the inner regions, where the arc structures seen in the
HST image are not detected in the IFS map. This is a combined effect of the
wider PSF and the worse sampling, that reduces the structural information in
the inner regions.  However, the A, B and C continuum dominated substructures
(see Fig.  \ref{hst}) are clearly identified in the IFS residual map.  This
comparison demonstrates that the applied technique is valid to recover the
substructures in this object. We overplotted the radio map at 4885 MHz, using
its WCS to align them with the HST and the IFS data. The radio data were taken
using the VLA in the A configuration, with a beam of
0.35$\arcsec\times$0.35$\arcsec$ \cite{walk97}.  There is no evident
connection between the continuum structures and the radio jet.
%\cite{soub89} proposed a scheme in which the A substructure
%is the result of a star formation process
%The 4$\arcsec$ knot is located over the A substructure, in
%projection, but its peak does not coincide with any of the different clumps
%detected in there. Evenmore, the A substructure extended further in the
%south-east than the knot, and therefore a starformation
We will discuss elsewhere about the nature of these structures (Garc\'\i
a-Lorenzo, in prep.).

Figure \ref{maps}, bottom-left panel, shows the contour-plot of the
narrow-band image centred at the emission line [OIII]$\lambda$5007, at the
redshift of the object (5170-5200\AA), extracted from the residual datacube.
The grey-scale shows the same narrow-band image of the continuum adjacent to
this line as shown in the top-right panel. The EELRs are now clearly
distiguished. There is no clear correspondence between the continuum-dominated
substructures and these emission-line dominated ones. 
%In particular, the
%substructure A does not correspond to the E$_1$ area, which shows a pure
%gaseous emission. 
The bottom-right panel shows the same contour-plot together with a
grey-scale representation of the radio map. The center of the E$_1$
region is located at $\sim$1$\arcsec$ north-west the radio knot at
4$\arcsec$ \citep{soub89}, just coincident with the bend in the radio
jet, which passes across the EELR. Despite of the projection effects,
this indicates most probably a physical connection between them
\citep{axon89,soub89}.
%\cite{axon89} reported the detection of an EELR
%associated with the undetected counter-jet
We cannot confirm the detection reported by \cite{axon89} of an EELR
associated with the undetected counter-jet, at $\sim$5$\arcsec$ east from the
nucleus.

\begin{figure}
  \includegraphics[angle=-90,width=\hsize]{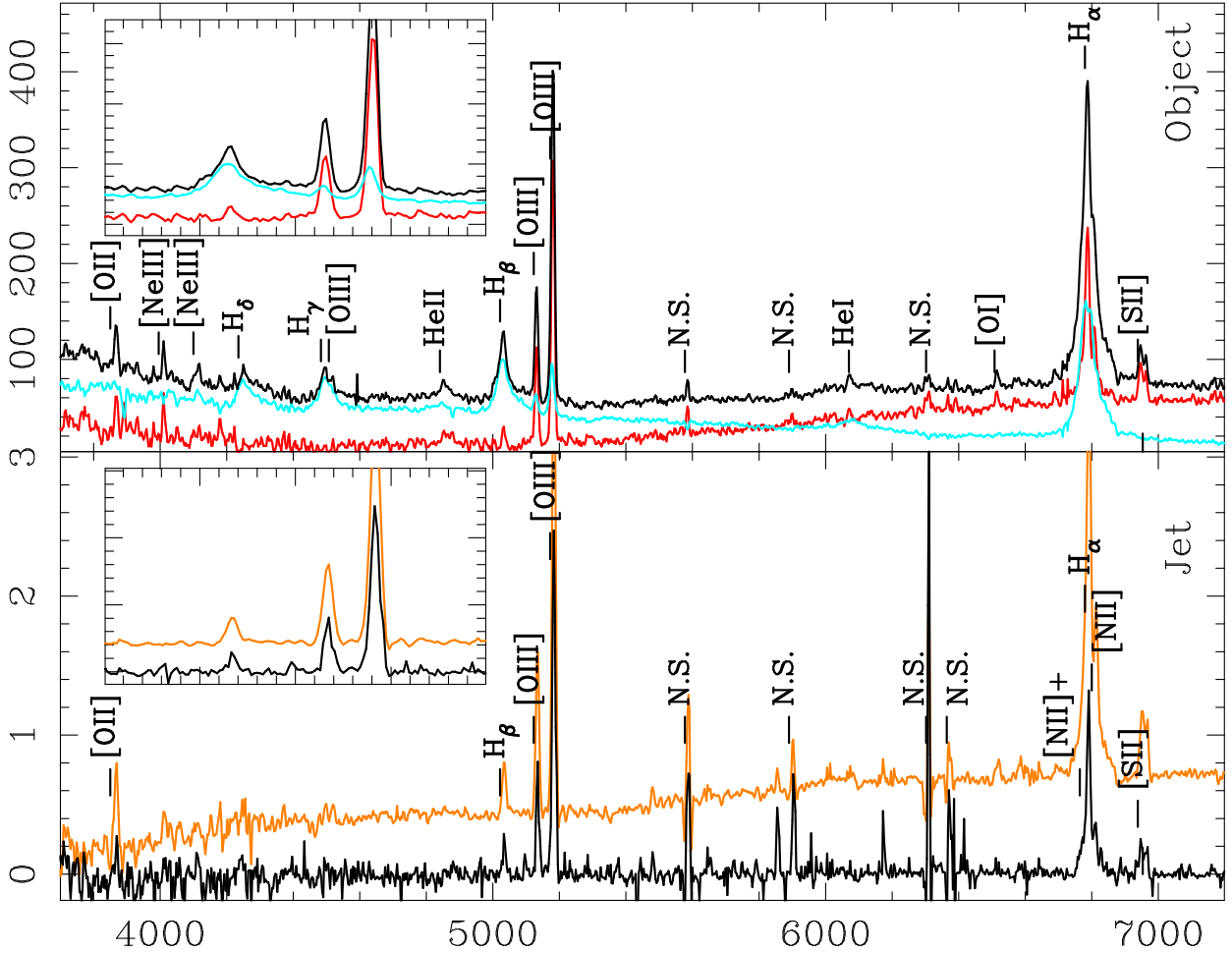}
  \caption{\label{spectra} 
    Top panel: Integrated spectrum of 3C 120 over the field of view of our
    observations in the wavelength range between 3700 and 7200 \AA~, in units
    of \fluxunits, with a detail over the [OIII]-H$\beta$ spectral range
    (small-box).  The blue and red-lines show the integrated spectrum of the
    nucleus and host galaxy, respectively, obtained from the 2D fitting.
    Bottom panel: Integrated spectrum of the E$_1$ area of the residual
    datacube after model subtraction in similar wavelength ranges. We have
    included the spectrum of the same area before the subtraction of the
    smooth component as an orange line. In both panels we have marked the
    detected emission lines.
  }
  \end{figure}

Figure \ref{spectra}, top panel, shows the integrated spectrum of
3C~120 in the wavelength range between 3700 and 7300 \AA, together
with the nucleus and host spectra obtained from the 2D fitting
technique. The detected emission lines have been indicated with their
corresponding names. A detail of the spectra in the
[OIII]$\lambda$5007/H$\beta$ spectral region is also presented.  The
spectrum of the nucleus is clearly bluer than the host galaxy one. It
contains all the broad emission lines, as expected. The narrow
emission lines are considerably fainter than the broad emission
lines. We only detected the [OIII] and [NII] lines (that blended with
the broad H$\alpha$).
%The average host galaxy spectrum is redder. 
%The 
The average host galaxy spectrum contains almost all the narrow
emission lines. Recent results indicate that the ionized gas producing
the observed narrow emission line in AGNs can extend throughout all
the host galaxy \citep{jahn02,jahn04}.  There is no clear 4000\AA\
break in the host spectrum, but a rise up in the blue-UV spectral
range. This indicates most probably a large amount of star formation
and/or a very young stellar population \citep{bald80,mole88}. However,
the line ratios, log$_{\rm 10}$([OIII]$\lambda$5007/H$\beta$)=1.63 and
log$_{\rm 10}$([NII]/H$\alpha$)=$-$0.39, indicate that the dominant
ionization source of the gas in this galaxy is the AGN. The average
dust content is high, with a Balmer ratio of
H$\alpha$/H$\beta$$\sim$10 ($A_V$$\sim$4 mag). We cannot confirm the
[OIII]$\lambda$5007/H$\beta$ values near to $\sim$1 in the nuclear
regions, reported by \cite{bald80} and \cite{hua88}. Taking an average
spectrum of the nucleus (1$\arcsec$ aperture radius), once
decontaminated from the broad emission line by a line-fitting
deblending, we find a ratio remarkably similar to the average over the
whole galaxy: log$_{\rm 10}$([OIII]$\lambda$5007/H$\beta$)=1.04.
%(log$_{\rm 10}$([NiII]/H$\alpha$)=$-$0.63).
%The Balmer ratio in this area is also
%high, but lower than the average (H$\alpha$/H$\beta$$\sim$13).

Figure \ref{spectra}, bottom panel, shows the {\it clean} spectrum of the
E$_1$ region, obtained by coadding the spectra of the residual datacube in a
1$\arcsec$ aperture centred in that region. For comparison, we plotted the
spectrum of this area before decontamination (i.e., a non-clean spectrum). The
differences between the spectra are clearly identified. The clean spectrum is
an emission line dominated spectrum, as expected from pure ionized gas, with
no significant continuum. On the other hand, the non-clean spectrum has a
significant continuum contribution and its emission lines are $\sim$2.4
brighter, which indicates a strong contamination from the host galaxy. The
Balmer ratio is H$\alpha$/H$\beta$$\sim$4 ($A_V$$\sim$1 mag) for the non-clean
spectrum, in contrast with the H$\alpha$/H$\beta$$\sim$7 ($A_V$$\sim$3 mag)
value for the clean one.  The [OIII]$\lambda$5007/H$\beta$ and [NII]/H$\alpha$
line ratios, $\sim 1.4$ and $\sim -0.4$, are similar in the clean and
non-clean spectra. They are also similar to the ratios found in the average
host-galaxy spectrum.

Based on the above quoted line ratios the main mechanisms which might be
involved in the emission line processes are: (1) photoionization by a hard UV
continuum, emitted most probably by the AGN, and (2) high velocity radiative
shocks which can influence the emission line processes due to the generation
of a strong local UV photon field in the host post-shock zone. Both processes
can generate the observed line ratios, under certain physical conditions
\citep{veil87,dopi95}. On the other hand, the line ratios exclude
photoionization from a star-forming region \citep{veil87}.
%\cite{vill03} shown that a good discriminator between
%both mechanisms is the position in the [OIII]$\lambda$5007/4363 versus
%HeII$\lambda$4686/H$_\beta$ diagram. However, we did not detect neither
%[OIII]$\lambda$4363 nor HeII$\lambda$4686 in the spectrum of E$_1$. Using the
%3$\sigma$ upper-limit for a line detection, we find that
%[OIII]$\lambda$5007/4363$>$11 and HeII$\lambda$4686/H$_\beta$

In the case of photoionization by the AGN, there is a simple relation between
the H$\beta$ and the nearby continuum luminosity (e.g., 4861\AA), assuming a
power-law for the ionizing continuum \citep{oste89}. Using the dust corrected
H$\beta$ luminosity, the ionizing continuum should have an intensity of $\sim
0.4\times$\fluxunits\ at 4861\AA. This intensity can be compared with the flux
received by the E$_1$ region from the AGN, using the spectrum of the nucleus
described above.  Assuming an isotropic emission and a $\sim 1/r^{2}$ decay of
the flux, and considering that the E$_1$ region has a diameter of $\sim$2~kpc
\citep{axon89}, and it is at a distance of $\sim$10~kpc from the nucleus, this
flux is $\sim 0.3\times$\fluxunits. Under these assumptions, the amount of UV
flux received by the E$_1$ region from the AGN would be enough to photoionize
it.

Assuming a photoionization mechanism we can derive the physical conditions in
the cloud \citep{oste89}. Using the upper-limit to the flux of the
undetected [OIII]$\lambda$4364 line, and the relation between the temperature
and the ([OIII]$\lambda$5007+4959)/[OIII]$\lambda$4363 flux ratio, we obtain
an upper limit to the effective temperature in the E$_1$ region of $<$55000 K.
We estimated the electron density, $n_e\sim$160~cm$^{-3}$, using the line
ratio [SII]$\lambda$6716/$\lambda$6731 ($\sim$1.3 for the clean spectrum) and
the relation between this ratio and the density (assuming a temperature of
$\sim$10$^{4}$K). These values are similar to those found in the EELRs
associated with jet-cloud interactions in other radio galaxies
\citep[e.g.][]{vill99,solo03}.

%This simple estimation 
%agreement with the flux required to produce the observed H$\beta$ luminosity.
%\cite{hjor95} measured the flux of the optical e
%Another source of ionization 

%As we quoted above a 
A post-shock zone can also give rise to the observed line
ratios in the case of high shock velocities, $\sim$350-500~km~s$^{-1}$, and
low magnetic fields, $B/\mu G\sim$0 cm$^{3/2}$ \citep{dopi95}. These
velocities are in the range of the lateral expansion velocity estimated by
\cite{axon89} for E$_1$, $\sim$350-700 km~s$^{-1}$, assuming a projection
angle $\sim$12-24\degr.  However, such a low magnetic field implies a density
lower than the previously derived from the [SII] line ratio.  \cite{dopi95}
already noticed that under the effects of a shock the density derived from the
[SII] line ratio is unreliable, due to the compression and the change of
ionization stage in different regions. For low magnetic fields this ratio can
be $\sim$1.3 for a pre-shock density of 1~cm$^{-3}$.  \cite{dopi96} determined
the relation between the H$\beta$ luminosity in the post-shock zone and the
shock velocity and electron density.  Using this relation it is required a
shock velocity of $\sim$340~km~s$^{-1}$ to reproduce the observed luminosity
of H$\beta$. This velocity is similar to the estimated lateral expansion speed
quoted above. Therefore, a shock process can also ionize the E$_1$ region.

\section{Conclusions}

We developed a new technique for decoupling the spectra of different
components in nearby galaxies and AGNs, using IFS. Using it we obtained, for
the first time, the decoupled spectra of the nucleus and the host galaxy of
3C~120, and a clean spectrum of the EELR associated with its radio-jet. Two
different mechanisms can cause the ionization of this region: direct
photoionization by the AGN or by UV photons emitted by the gas cooling behind
a shock front.  In both cases, there are compelling evidences that the
jet-cloud interaction plays a major role: (a) The lateral expansion of the jet
and its interaction with the ISM is, most probably, the reason for the density
enhancement in the E$_1$ area \citep{axon89}; (b) In the case of direct
photoionization the UV flux needs to reach the region without substantial
absorption.  The jet itself, which has associated high-energetic particles,
can destroy the dust grains creating the observed dust decrease in this
region; (c) In the case of a shock ionization, the lateral expansion of the
jet produces the shock itself.
%Only
%in a few cases it was possible to distinguish between both mechanisms
%\citep[e.g.][]{vill99}. 
Maybe the combined effect of a direct AGN and a shock induced photoionization
has to be considered in order to understand the ionization of the E$_1$
region.% of 3C~120.

\section{Acknowledgments}

This project is part of the Euro3D RTN on IFS, funded by the EC under contract
No.  HPRN-CT-2002-00305. The WHT is operated on the island of La Palma by the
Isaac Newton Group in the Spanish Observatorio del Roque de los Muchachos of
the IAC.  This project has used images obtained from the HST archive, using
the ESO archiving facilities.  We would like to thank Dr.Walker that has
kindly provided us with the radio maps of 3C~120.  We would like to thank the
anonymous referee that has helped us to improve the quality of this paper
with his/her remarks.

%\bibliography{../gems/low_z_agns/jahns}

% \bibliography{3c120}
 \label{lastpage}

 %\newpage

\end{document}